\documentclass[12pt,amsmath,amssymb,twocolumn]{iopart}

\usepackage{epsfig}

\begin{document}

\title{Effect of diffusion in simple discontinuous absorbing transition models}

\author{Salete Pianegonda}
\address{Instituto de F\'{\i}sica, Universidade Federal 
do Rio Grande do Sul, Caixa Postal 15051, CEP 91501-970, 
Porto Alegre, RS, Brazil}
   
\author{Carlos E. Fiore}
\address{Instituto de F\'{\i}sica,
Universidade de S\~{a}o Paulo, 
Caixa Postal 66318\\       
05315-970 S\~{a}o Paulo, SP, Brazil}
\date{\today}

\begin{abstract}
Discontinuous transitions into absorbing states require an effective
mechanism that prevents the stabilization of low density states.
They can be found in different systems, such as lattice models
or stochastic differential equations (e.g.  Langevin equations).
Recent results for the latter approach have shown that 
 the inclusion of limited diffusion  suppresses discontinuous transitions, 
whereas they are maintained for larger diffusion strengths. Here we give a
further step by addressing the effect of diffusion in two simple lattice
models originally presenting discontinuous absorbing transitions.
They have been studied via mean-field theory  (MFT)
and distinct sort of numerical simulations. For both cases,
results suggest that  
the diffusion does not change the order of the transition, regardless 
its strength and thus, in 
partial contrast with results obtained from Langevin approach. Also, 
all  transitions present a common finite size scaling behavior that 
is similar to discontinuous absorbing transitions
 studied  in Phys. Rev. E {\bf 89}, 022104 (2014).
\end{abstract}

\maketitle

\section{Introduction}

Nonequilibrium phase transitions into absorbing states  describe 
several problems, such as wetting phenomena,
spreading of diseases, chemical reactions and others \cite{r1,r2}.
In the last years, a large effort for its characterization
including experimental
verifications \cite{r3,r3.1,r3.2} and 
the establishment of distinct universality
classes \cite{r1,r2,chate} have been undertaken.
Generically, continuous phase transitions into 
an absorbing state for systems without conservation laws nor
extra symmetries fall into in the directed percolation (DP) universality 
class \cite{r2,r4,r5}. 
The best example of this category 
is probably the contact process (CP) \cite{r4}. It is defined
on a given $d$-dimensional lattice and the dynamics comprehends the
 creation in the presence of at least one adjacent
particle and spontaneous annihilation. Since the particle creation
requires the presence of adjacent particles, a configuration
devoided of species is absorbing.

Conversely, discontinuous absorbing
transitions also appear in different systems
 \cite{zgb,bid,r7,grass,r9}, but  comparatively 
 they have  received less attention than the continuous ones.
Recently, they have attracted  interest  for the search of
minimal ingredients for their occurrence, being
one possibility  the so called ``restrictive'' (threshold) contact processes (CPs).
These models are variants of the second Sch\"ogl model, in which  
the particle creation is similar to the usual CP, but 
instead it requires a minimal
neighborhood larger than 1 particle for creating a new species
\cite{evans,evans2,oliveira2,oliveira,carlos}. 
Different studies have stated that the phase transition,
induced  by this mild  change (with
respect to the usual CP), remains unaffected 
under  the inclusion of   distinct creation 
\cite{evans,evans2,oliveira2,oliveira,carlos} and 
annihilation rules \cite{carlos}
as well as  for different lattice topologies  \cite{durret}.
In some specific cases \cite{evans,evans2}, in which 
the particle creation occurs only if one has at least two 
adjacent diagonal pairs
of particles, the phase transition is characterized by a generic two-phase
coexistence and exhibits an interface orientational dependence at the transition point.  On the other hand, when the transition rates  depend only
 on the number of nearest neighbors particles 
(and not their orientations)   the discontinuous transitions 
take place at  a single point \cite{oliveira2, oliveira,carlos}.

Despite the  apparent robustness of first-order transitions
for the above mentioned restrictive examples, 
the effect of some (relevant) dynamics
has been so far unexplored in the present context. These 
dynamics, such as  spatial disorder
and particle diffusion,  can cause drastic changes in  
continuous phase transitions \cite{r2}.  One of 
the few available studies shows that spatial disorder
suppresses the phase coexistence, giving rise   
to a continuous transition belonging to a new 
universality class \cite{munoz14}.
A similar scenario of scarce results   also
holds for the outcome of diffusion.
Very recently, a stochastic differential equation  
(such as a Langevin equation)
  reported that different diffusion strengths 
 lead to opposite findings (in two dimensions). 
Whenever  the transition 
is  discontinuous for larger diffusion values, limited  rates suppress it, 
giving rise to a critical phase transition belonging to the DP
universality class \cite{munoz214}.
With these ideas in mind, we give a further step by tackling the 
influence of diffusion in  lattice systems presenting 
discontinuous absorbing phase transitions.  
\cite{oliveira2,oliveira}. Our study aims to answer three 
fundamental points: (i) what is the effect of strong and 
limited diffusion in these cases? (ii) does it suppress 
the discontinuous transition? (iii) How does our results compare with those
obtained from the coarse-grained description in  Ref. \cite{munoz214}? 
In other words, are there 
differences between lattice models and Langevin equations?

We consider two lattice models and  three
representative values of diffusion,  in order to exemplify the 
low, intermediate and large regimes. 
Models will be studied via MFT  and distinct kinds of numerical simulations 
(explained further).
Results suggest that the  discontinuous transition 
is maintained in both models for all diffusion values.
Also, a finite-size scaling behavior similar to
discontinuous transitions studied in Ref. \cite{carlos} has been found.
 Since there is no  
theory for the nonequilibrium case,
our results can shed  light over a general finite-size scaling for
 them.\\
This paper is organized as follows: Sec. II presents the models and mean-field
analysis. Sec. III shows the numerical results
and finally conclusions are drawn 
in Sec. IV.

\section{Models  and mean-field analysis}

Let us consider  systems of interacting particles placed on a square lattice
of linear size $L$.  Each site has an occupation  
variable $\eta_i$ 
that assumes the value $0\,\,(1)$ whenever sites are empty (occupied). 
The  model A is defined by the following interaction rules: particles are
 annihilated with rate $\alpha$ and are
 created   in empty sites only if their number of nearest neighbor
particles $nn$ is larger than 1 ($nn \ge 2$), with rate $nn/4$ \cite{oliveira2}. 
There is no particle
creation if $nn \le 1$. The model B is similar to the model A, but 
the particle creation rate always reads $1$, provided  $nn \ge 2$ \cite{oliveira}.
Thus, whenever 
particles are created with rate proportional to the
number of their nearest neighbors for the model A, it is independent on $nn$
in the model B. Besides
the above creation-annihilation dynamics, each particle also
hops  to one of its nearest neighbor sites
with probability $D$, provided it is empty. 
In the regime of low annihilation parameters, the system
exhibits indefinite activity in which particles are 
continuously created and destroyed.
In contrast, for larger $\alpha$'s, the system is constrained
in the absorbing phase. The phase transition 
separates above regimes at a transition point $\alpha_{0}$. 
In the absence of diffusion, the phase transitions for both
models A and B 
are discontinuous and occur at 
$\alpha_0=0.2007(6)$ and $0.352(1)$,
respectively \cite{oliveira2,oliveira}.

The first inspection of the effect of diffusion
can be achieved by performing  mean-field analysis. The starting
point is to write down  the time evolution of relevant quantities
from the interaction rules and truncating the associated probabilities
at a given level. 
Since the diffusion conserves the  number of particles, 
it is required to take into account at least correlations
of two sites   and 
hence two equations are needed. Designating 
the symbols $\bullet$  and $\circ$  to
represent occupied ($\eta_i=1$) and ($\eta_i=0$) empty sites, 
the system density $\rho$ corresponds to the one-site probability 
$\rho=P(\scalebox{0.85}{$\bullet$})$. Another quantity to
be considered here is  
the two-site correlation  given by
$u=P(\scalebox{0.85}{$\circ$}\scalebox{0.85}{$\bullet$})$. 
From the above model rules, it follows
that 
\begin{equation}
\frac{d\rho}{dt}=2P(\scalebox{0.85}{$\circ$}\scalebox{0.85}{$\bullet$}\scalebox{0.85}{$\bullet$}
\scalebox{0.85}{$\circ$}
\scalebox{0.85}{$\circ$})+P(\scalebox{0.85}{$\circ$}\scalebox{0.85}{$\bullet$}\scalebox{0.85}{$\circ$}
\scalebox{0.85}{$\bullet$}
\scalebox{0.85}{$\circ$})+3P(\scalebox{0.85}{$\circ$}\scalebox{0.85}{$\bullet$}\scalebox{0.85}{$\circ$}
\scalebox{0.85}{$\bullet$}
\scalebox{0.85}{$\bullet$})+P(\scalebox{0.85}{$\circ$}\scalebox{0.85}{$\bullet$}\scalebox{0.85}{$\bullet$}
\scalebox{0.85}{$\bullet$}
\scalebox{0.85}{$\bullet$})-\alpha P(\scalebox{0.85}{$\bullet$}),
\end{equation}
and
\begin{eqnarray}
\label{D1}
\frac{du}{dt}&=&(1-D)[-\frac{3}{2}P(\scalebox{0.85}{$\circ$}\scalebox{0.85}{$\bullet$}\scalebox{0.85}{$\circ$}
\scalebox{0.85}{$\bullet$}
\scalebox{0.85}{$\bullet$})-P(\scalebox{0.85}{$\circ$}\scalebox{0.85}{$\bullet$}\scalebox{0.85}{$\bullet$}
\scalebox{0.85}{$\bullet$}\scalebox{0.85}{$\bullet$})\\[0.2cm]\nonumber
&&-\alpha\,P(\scalebox{0.85}{$\circ$}\scalebox{0.85}{$\bullet$})
+\alpha\,P(\scalebox{0.85}{$\bullet$}\scalebox{0.85}{$\bullet$})]
+D[6P(\scalebox{0.85}{$\bullet$}\scalebox{0.85}{$\circ$}
\scalebox{0.85}{$\bullet$})-6P(\scalebox{0.85}{$\circ$}\scalebox{0.85}{$\bullet$}
\scalebox{0.85}{$\bullet$})],\nonumber
\end{eqnarray}
for the model A 
 and
\begin{equation}
\frac{d\rho}{dt}=4P(\scalebox{0.85}{$\circ$}\scalebox{0.85}{$\bullet$}\scalebox{0.85}{$\bullet$}
\scalebox{0.85}{$\circ$}
\scalebox{0.85}{$\circ$})+2P(\scalebox{0.85}{$\circ$}\scalebox{0.85}{$\bullet$}\scalebox{0.85}{$\circ$}
\scalebox{0.85}{$\bullet$}
\scalebox{0.85}{$\circ$})+4P(\scalebox{0.85}{$\circ$}\scalebox{0.85}{$\bullet$}\scalebox{0.85}{$\circ$}
\scalebox{0.85}{$\bullet$}
\scalebox{0.85}{$\bullet$})+P(\scalebox{0.85}{$\circ$}\scalebox{0.85}{$\bullet$}\scalebox{0.85}{$\bullet$}
\scalebox{0.85}{$\bullet$}
\scalebox{0.85}{$\bullet$})-\alpha P(\scalebox{0.85}{$\bullet$}),
\end{equation}

\begin{eqnarray}
\label{D2}
\frac{du}{dt}&=&(1-D)[-2P(\scalebox{0.85}{$\circ$}\scalebox{0.85}{$\bullet$}\scalebox{0.85}{$\circ$}
\scalebox{0.85}{$\bullet$}
\scalebox{0.85}{$\bullet$})-P(\scalebox{0.85}{$\circ$}\scalebox{0.85}{$\bullet$}\scalebox{0.85}{$\bullet$}
\scalebox{0.85}{$\bullet$}\scalebox{0.85}{$\bullet$})\\[0.2cm]\nonumber
&&-\alpha\,P(\scalebox{0.85}{$\circ$}\scalebox{0.85}{$\bullet$})+\alpha\,P(\scalebox{0.85}{$\bullet$}\scalebox{0.85}{$\bullet$})]
+D[6P(\scalebox{0.85}{$\bullet$}\scalebox{0.85}{$\circ$}
\scalebox{0.85}{$\bullet$})-6P(\scalebox{0.85}{$\circ$}\scalebox{0.85}{$\bullet$}
\scalebox{0.85}{$\bullet$})],\nonumber
\end{eqnarray}
for the model B.
Here the symbol $P(\eta_0,\eta_1,\eta_2,\eta_3,\eta_4)$  denotes the probability
of finding the central site in the state $\eta_0$ and its four nearest
neighbors in the states $\eta_1$, $\eta_2$, $\eta_3$ and $\eta_4$.

The pair mean-field approximation consists of rewriting
the $n$-site probabilities ($n>2$) as
products of two-site in such a way that
\begin{equation}\label{pair_approx}
P(\eta_0,\eta_1,...,\eta_{n-1}) \simeq \frac{P(\eta_0,\eta_1) P(\eta_0,\eta_2)...P(\eta_0,\eta_{n-1})}{P(\eta_0)^{n-2}}\,.
\end{equation}
From this approximation,  the above equations read
\begin{equation}
\frac{d\rho}{dt}=\frac{3u^{2}}{(1-\rho)}-\frac{3u^{3}}{(1-\rho)^{2}}+
\frac{u^{4}}{(1-\rho)^{3}}-\alpha\rho,
\end{equation}
\begin{eqnarray}
\frac{du}{dt}&=&(1-D)\left[-\frac{3u^{3}}{2(1-\rho)^{2}}+\frac{u^{4}}{2(1-\rho)^{3}}
-2\alpha\,u+\alpha\rho\right]+\nonumber\\ [0.2cm]
&&+D\left[6u-\frac{6u^{2}}{\rho(1-\rho)}\right],
\end{eqnarray}
for the model A 
and
\begin{equation}
\frac{d\rho}{dt}=\frac{6u^{2}}{(1-\rho)}-\frac{8u^{3}}{(1-\rho)^{2}}+
\frac{3u^{4}}{(1-\rho)^{3}}-\alpha\rho,
\end{equation}
\begin{eqnarray}
\frac{du}{dt}&=&(1-D)\left[-\frac{2u^{3}}{(1-\rho)^{2}}+\frac{u^{4}}{(1-\rho)^{3}}
-2\alpha\,u+\alpha\rho\right]+\nonumber\\ [0.2cm]
&&+D\left[6u-\frac{6u^{2}}{\rho(1-\rho)}\right],
\end{eqnarray}
for the model B.

The steady solutions are obtained 
by taking $\frac{d\rho}{dt}=\frac{du}{dt}=0$ implying that
for locating
the transition point and the order 
of transition it is required to solve a system of two coupled equations 
for a given set of parameters ($\alpha$,\,$D$).  
Although alternative treatments have 
been considered  \cite{evans3}, here we shall identify the
order of transitions by  inspecting the dependence
of $\rho$ vs $\alpha$. In similarity with equilibrium
transitions,  the existence of a spinodal
behavior (with $\rho$ increasing by raising $\alpha$) signals a 
discontinuous transition. Despite this, no 
analogous treatments similar to the ``Maxwell construction'' are
available.  For instance, the coexistence points 
have been estimated by the maximum value  of $\alpha$
and the spinodal behavior replaced by a jump in $\rho$.

Fig. 1 $(a)$ and $(b)$ show the phase diagram for distinct diffusion rates. 
For all values of $D$, mean-field results predict  discontinuous transitions 
separating the absorbing 
and the active phases. However, as $D$
increases, the transition point moves to larger values of $\alpha_0$
and the active phase  becomes less dense.
For example, for $D=0.1$ and $D=0.9$, the active phases 
have densities $\rho_{ac}=0.445$ and $0.371$ (model A) and
 $\rho_{ac}=0.419$ and $0.324$ (model B), respectively. 
For $D \rightarrow 1$, one recovers the limit 
of fully  uncorrelated particles, so that
$P(\eta_0,\eta_1,...,\eta_{n-1})=P(\eta_0) P(\eta_1)             
...P(\eta_{n-1})$. In this limit,
Eqs. (1) and (3) become equivalent to those
obtained from the one-site MFT given by 
$\frac{d \rho}{dt}=\rho^2(1-\rho)[3-3\rho+\rho^2]-\alpha \rho$
and $\frac{d \rho}{dt}=\rho^2(1-\rho)[6-8\rho+3\rho^2]-\alpha \rho$, for the models A and B,
respectively. For such regime,  
 transition points take  place at $\alpha_0=0.4724$ 
with active phase density $\rho_{ac}=0.370$ (model A) and
$\alpha_0=0.8154$ with  $\rho_{ac}=0.322$ (model B).
Thus, despite the increase of diffusion displaces particles, 
MFT predicts a discontinuous transition,
regardless the strength of the diffusion rate. As a difference 
between models, active phases
are somewhat more dense for  model A than in model B.

\begin{figure}[h]
\centering
\includegraphics[scale=0.37]{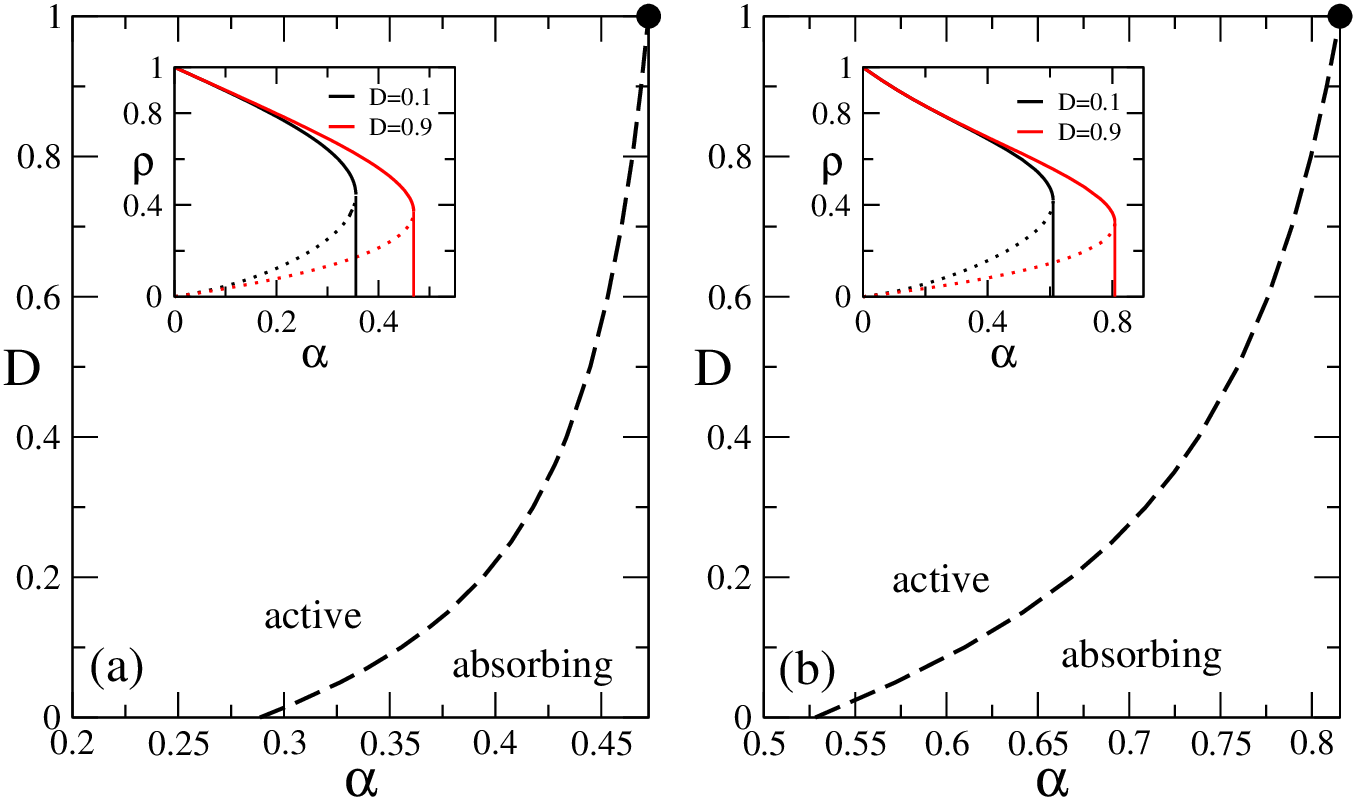}
\caption{Two-site mean field phase diagrams for diffusive model versions
A $(a)$ and B $(b)$. Dashed  lines
denote discontinuous  phase transitions. The black
circles  indicate the  $D \rightarrow 1$ limit predicted by the one site MFT.
Insets show  the two-site MFT results for $D=0.1$ and $D=0.9$. Dotted
lines correspond to spinodal behaviors, being  replaced here by a jump.}
\label{fig1}
\end{figure}

\section{Numerical results}
Numerical simulations have been performed for   distinct system sizes
of a square lattice with  periodic boundary conditions.
Due to the lack of a  general theory   for discontinuous
absorbing transitions, including  the absence of a finite-size scaling (FSS) 
theory and unknown spreading experiments behavior for $d \ge 2$, 
two sort of analysis will be presented.  
In the former, we study the time decay of  the density $\rho$
starting from a fully occupied initial condition for distinct independent runs.
As for critical 
and discontinuous phase transitions, for small
$\alpha$ the density $\rho$ converges to a
 definite value indicating endless activity,
in which particles are continuously 
created and annihilated. On the contrary, for 
sufficiently large $\alpha$'s, the system density  $\rho$ vanishes
exponentially toward a complete particle extinction. 
The ``coexistence'' point 
$\alpha_{0}$ is the separatrix between above regimes, whereas
at the critical point $\alpha_c$ the density $\rho$ vanishes algebraically 
following a power-law behavior $\rho \sim t^{-\theta}$,  with $\theta$ its 
associated critical exponent. 
For the DP universality class it reads $\theta=0.4505(10)$ \cite{henkel}.
Thus, the difference of
above behaviors will be used to identify the order of phase transition.

Second, the behavior of typical quantities in the steady 
regime is investigated. For instance, we apply
 the models dynamics together with the quasi-steady method
\cite{qs}. Briefly, it consists
of  storing a list of $M$ active configurations (here we
store $M=2000-3000$ configurations)
and whenever the system falls into the absorbing state a
configuration is  randomly extracted from the list.
The ensemble of
stored configurations is continuously updated, where
for each MC step a configuration belonging to the list is replaced with
probability ${\tilde p}$ (typically one takes ${\tilde p}=0.01$)
by the actual system configuration, provided it is not absorbing.
Discontinuous transitions are typically signed by
bimodal probability distribution $P_\rho$ (characterizing
the ``coexistence'' between absorbing and active phases) and 
for finite systems, a peak in the 
order-parameter variance 
$\chi=L^{2}[\langle \rho^{2}\rangle-\langle \rho \rangle^{2}]$ is expected
 close to the transition point.
For equilibrium systems, the maximum of $\chi$ and other quantities scale
with the system volume and its position $\alpha_L$
obeys the asymptotic relation $\alpha_L =\alpha_0 - c/L^2$ \cite{rBoKo},
being $\alpha_0$ the transition point in the
thermodynamic limit and $c$ a constant. 
Although the finite size scaling properties are unknown for nonequilibrium systems, results
for some first-order transitions into an absorbing 
state have shown a similar scaling 
than the equilibrium case (with the system volume) \cite{carlos,sina}.

\begin{figure}[h]
\centering
\includegraphics[scale=0.35]{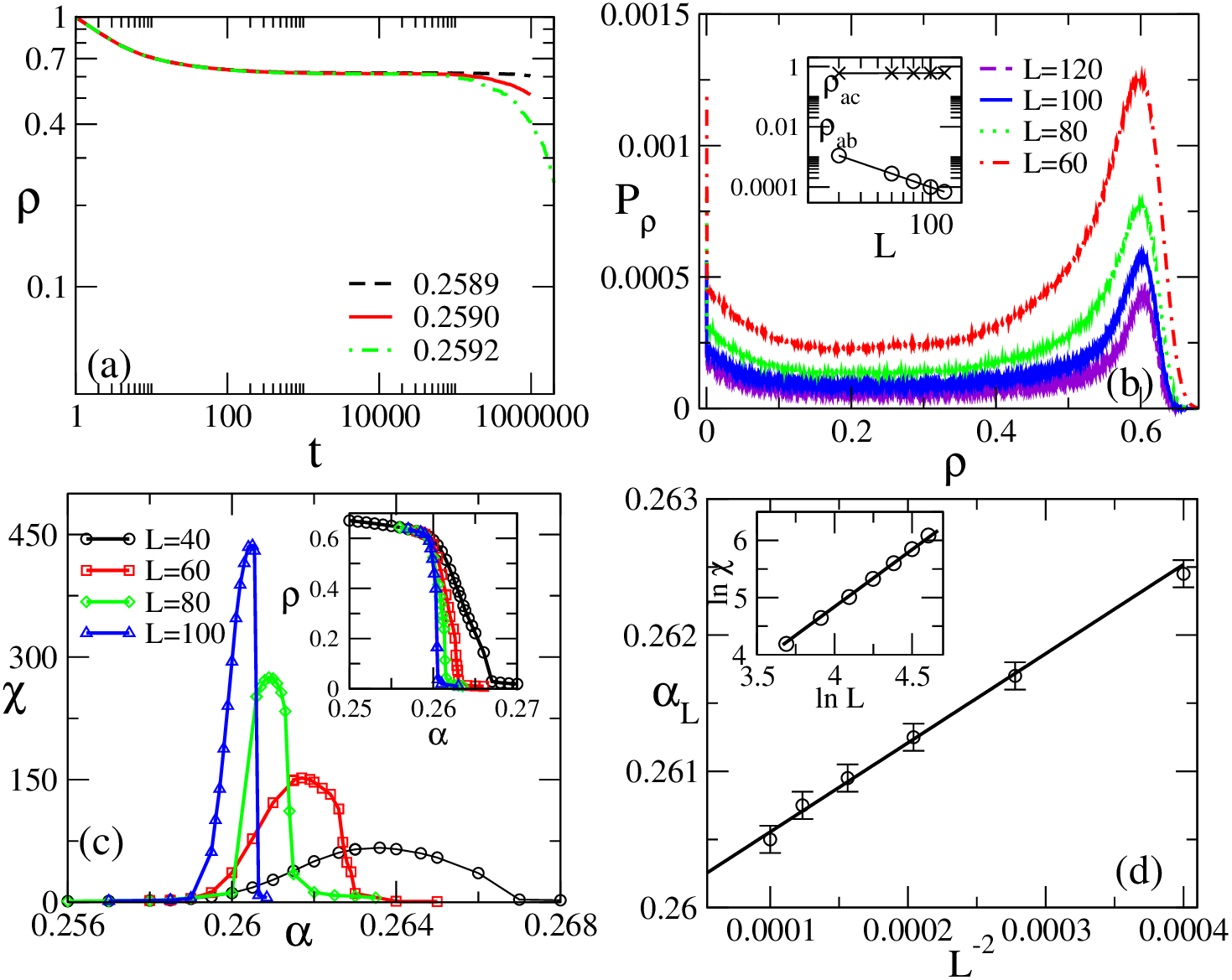}
\caption{For the model A and $D=0.1$,  panel $(a)$ shows the time evolution of $\rho$ 
for distinct $\alpha$'s and $L=150$. In $(b)$ the quasi-stationary probability distribution              
$P_{\rho}$  for $L$'s  in which the peaks
have the same height. In the inset, the  log-log plot 
of the quasi-steady densities vs L. In $(c)$
$\chi$ and $\rho$ (inset) vs $\alpha$ 
for distinct system sizes $L$. In $(d)$, the scaling plot
of $\alpha_{L}$, in which $\chi$ is maximum, vs $L^{-2}$. Inset shows the
log-log plot of the maximum of $\chi$ vs $L$ and the straight line has
slope $2$.}
\label{fig2}
\end{figure}

\begin{figure}[h]
\centering
\includegraphics[scale=0.35]{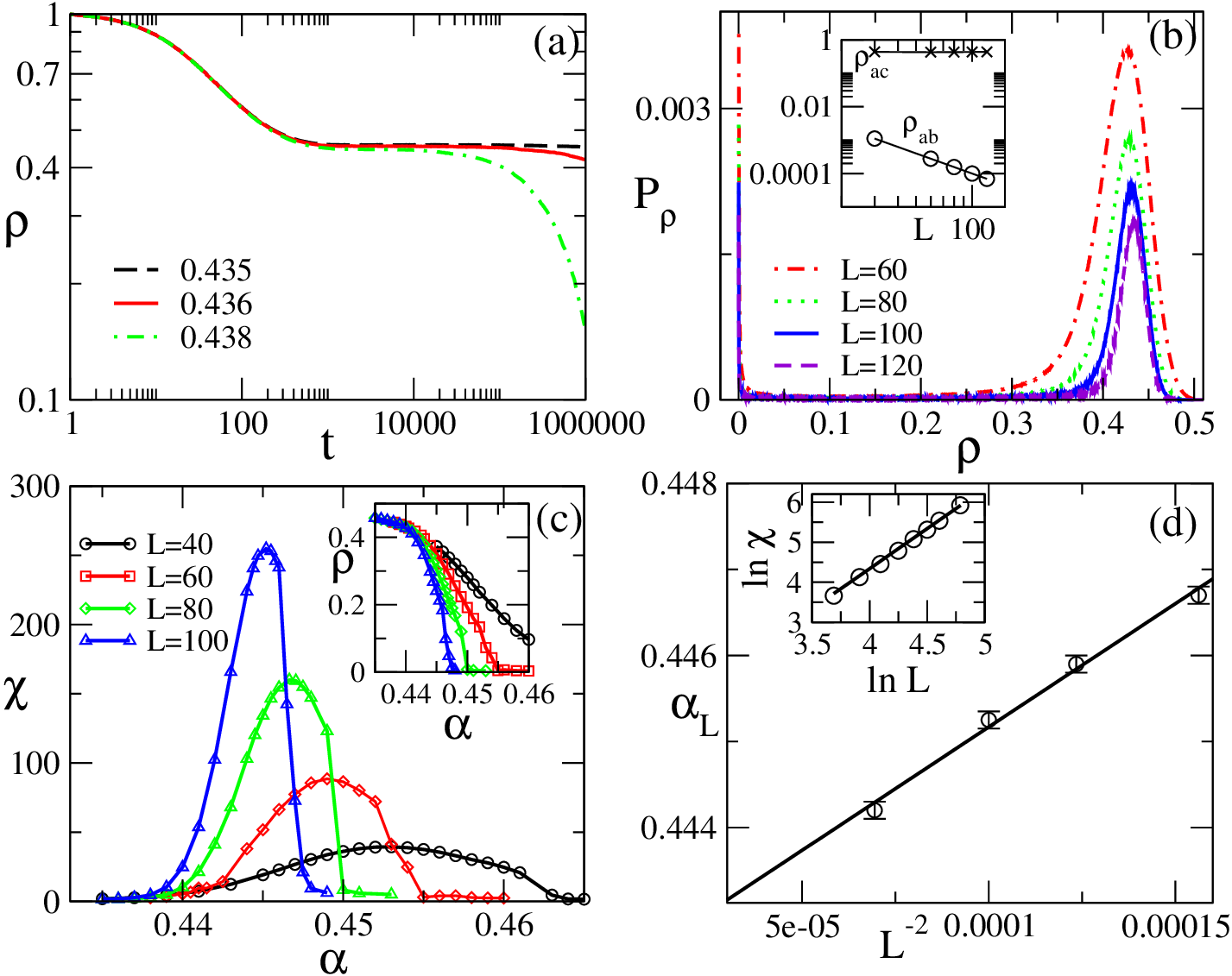}
\caption{For the model A and $D=0.9$, panel $(a)$ shows the time evolution of $\rho$ 
for distinct $\alpha$'s and $L=150$. In $(b)$ the quasi-stationary probability distribution              
$P_{\rho}$  for $L$'s  in which the peaks
have the same height. In the inset, the  log-log plot 
of the quasi-steady densities vs L. In $(c)$
$\chi$ and $\rho$ (inset) vs $\alpha$ 
for distinct system sizes $L$. In $(d)$, the scaling plot
of $\alpha_{L}$, in which $\chi$ is maximum, vs $L^{-2}$. Inset shows the
log-log plot of the maximum of $\chi$ vs $L$ and the straight line has
slope $2$.}
\label{fig4}
\end{figure}

Results for the model A and three representative 
diffusion rates ($D=0.1$, $D=0.5$ (not shown)  
and $0.9$) are summarized in Figs. \ref{fig2} and \ref{fig4}.
In all cases, panels $(a)$ show that there is a 
threshold point $\alpha_0$ separating an active state 
(signed by the convergence to a definite value of $\rho$) 
from an exponential vanishing of $\rho$.
Such values increase by raising $D$, yielding at  
$\alpha_0 \sim 0.2590$, $0.360$ and $0.436$ for $D=0.1$, $0.5$ and $0.9$, respectively.
Since no  power-law behavior is presented, such analysis 
 provides the first evidence of 
a discontinuous transition for all diffusion strengths. 
In order to confirm this, panels $(b)$ show the probability distribution $P_{\rho}$
for distinct system sizes. In all cases $P_{\rho}$ presents a bimodal shape, whose
position of equal peaks deviate mildly as $L$ increases. Also, the peaks
corresponding to the active and absorbing phases present distinct dependencies on $L$.
Whereas active phase densities $\rho_{ac}$ converge to 
well defined values,  $\rho_{ab}$ vanishes with $1/L^2$ (insets). For
$D=0.1,0.5$ and $0.9$ the $\rho_{ac}$'s  converge to
$0.603(1)$, $0.501(2)$ and $0.434(2)$, respectively. These features
are similar to the  restrictive models studied in Ref. \cite{carlos}.
For $D=0.99$ (not shown) a bimodal probability distribution
with active phase centered at $\rho_{ac}=0.405(1)$ is observed.
In the third analysis, the behavior of
 the system density $\rho$ and its variance $\chi$ is presented.
Note that $\rho$ vanishes in a short range of $\alpha$ followed
by a peak of the variance $\chi$ (panels $(c)$ and their insets). 
For each $D$, the positions $\alpha_L$'s, in which  $\chi$ presents a maximum,
scale with $L^{-2}$  whereas the maximum of $\chi$ increases with $L^2$
(panels $(d)$ and  insets). These features are similar 
to equilibrium   discontinuous phase transitions \cite{rBoKo,fioreprl}
and with the scalings  in Ref. \cite{carlos}. 
From this behavior,  we obtain
the values  $\alpha_0=0.2600(1)$ ($D=0.1$), $0.3624(2)$ ($D=0.5$) 
and $0.442(1)$ ($D=0.9$), which  agree with previous estimates, obtained
from the time decay of $\rho$.  Small discrepancies between
estimates can be attributed to the lattice simulated
for the time decay be finite,  due to the 
uncertainties in the position of peaks or both. Thus, steady analysis 
reinforces
above conclusions concerning the phase transition be discontinuous
regardless the diffusion rate.


\begin{figure}[h]
\centering
\includegraphics[scale=0.35]{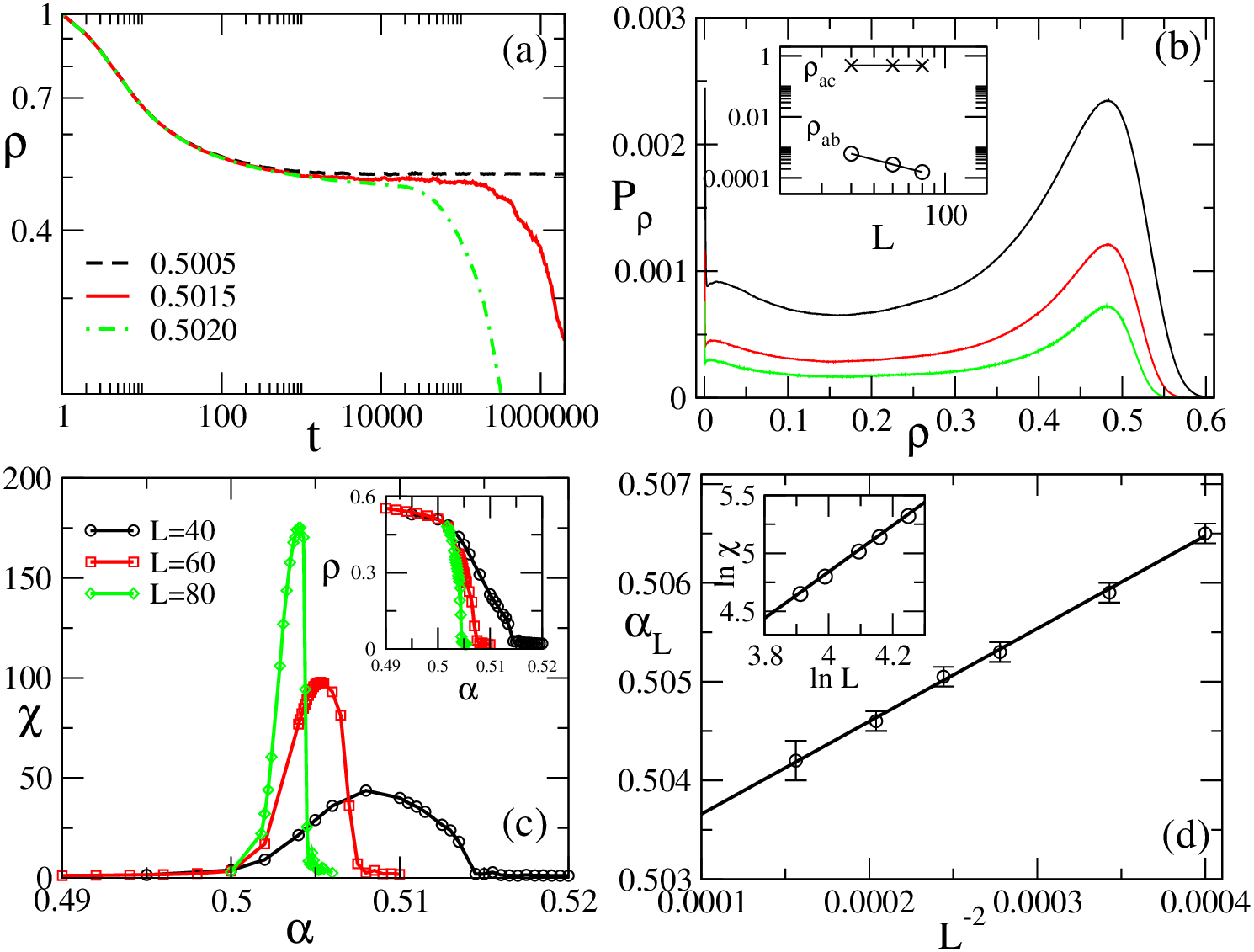}
\caption{For the model B and $D=0.1$,
panel $(a)$ shows the time evolution of $\rho$ 
for distinct $\alpha$'s and $L=150$. In $(b)$ the quasi-stationary probability distribution              
$P_{\rho}$  for $L$'s  in which the peaks
have the same height. In the inset, the  log-log plot 
of the quasi-steady densities vs L. In $(c)$
$\chi$ and $\rho$ (inset) vs $\alpha$ 
for distinct system sizes $L$. In $(d)$, the scaling plot
of $\alpha_{L}$, in which $\chi$ is maximum, vs $L^{-2}$. Inset shows the
log-log plot of the maximum of $\chi$ vs $L$ and the straight line has
slope $2$.}
\label{fig42}
\end{figure}

\begin{figure}[h]
\centering
\includegraphics[scale=0.345]{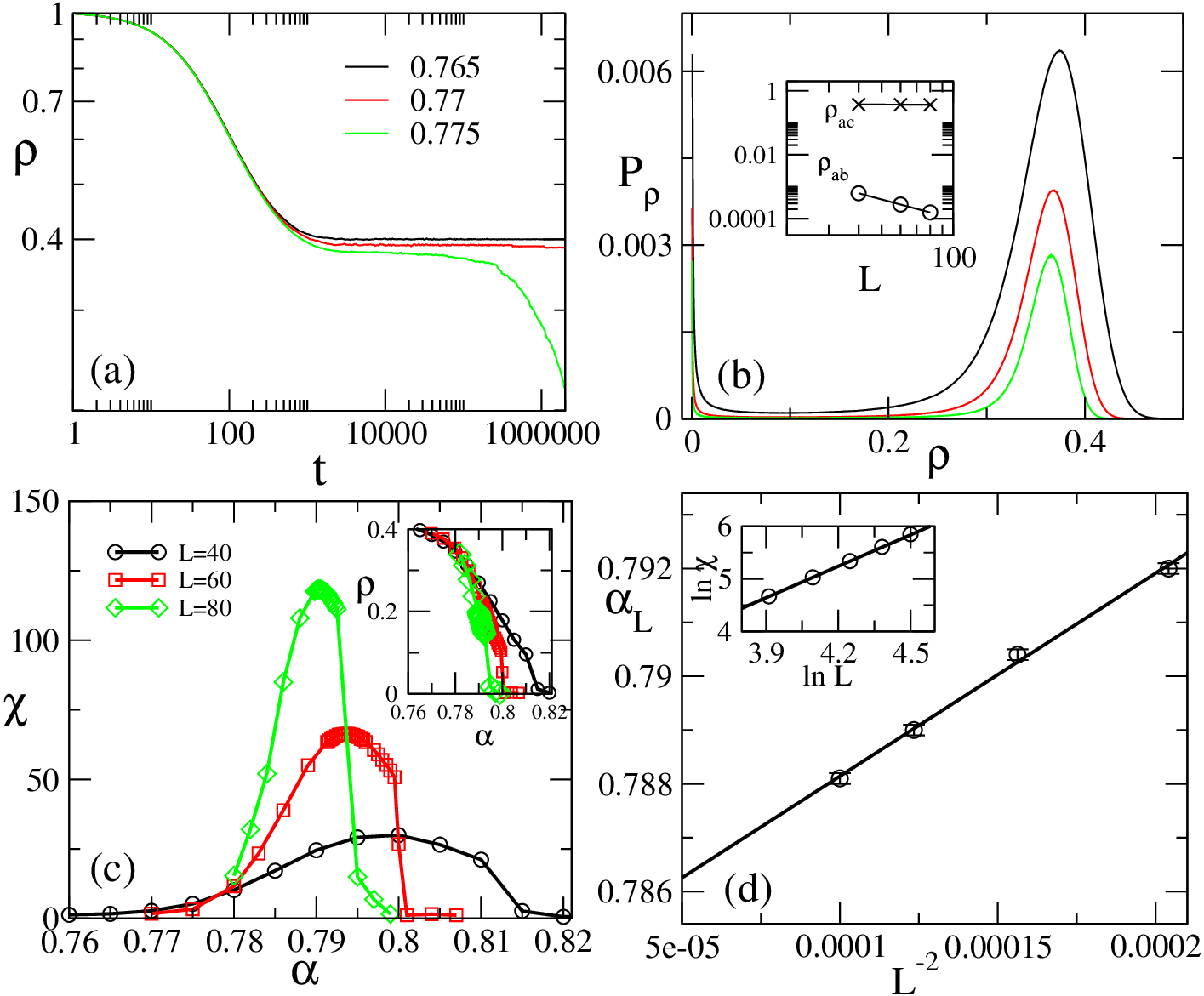}
\caption{For the model B and $D=0.9$,  panel $(a)$ shows the time evolution of $\rho$ 
for distinct $\alpha$'s and $L=150$. In $(b)$ the quasi-stationary probability distribution              
$P_{\rho}$  for $L$'s  in which the peaks
have the same height. In the inset, the  log-log plot 
of the quasi-steady densities vs L. In $(c)$
$\chi$ and $\rho$ (inset) vs $\alpha$ 
for distinct system sizes $L$. In $(d)$, the scaling plot
of $\alpha_{L}$, in which $\chi$ is maximum, vs $L^{-2}$. Inset shows the
log-log plot of the maximum of $\chi$ vs $L$ and the straight line has
slope $2$.}
\label{fig43}
\end{figure}

Next, we extended above analysis for  model B, 
with results summarized in Figs. \ref{fig42} and \ref{fig43}
for $D=0.1$ and $D=0.9$, respectively. As for the model A, both values
of $D$ show that the  separatrix between active and absorbing
regimes are signed by the absence of a power-law behavior and thus the 
transitions seem to be discontinuous.
In particular, the separatrix points (panels $(a)$) yield close to 
$\alpha_0 \sim 0.5015$ ($D=0.1$) and $0.770$ 
($D=0.9$).
In order to confirm, we also examine the probability distribution
as well as the behavior of $\rho$ and $\chi$.
In the former, $P_{\rho}$  also presents bimodal shapes exhibiting two well 
defined peaks (panels $(b)$). Whenever $\rho_{ac}$'s (insets) 
saturate as $L$  increases, the $\rho_{ab}$'s vanish as $1/L^2$. 
In addition,  active
phase also becomes less dense as $D$ increases, reading $\rho_{ac}=0.482(1)$ 
and $0.372(1)$
for $D=0.1$ and $0.9$, respectively.
As for the model A and those studied in Ref. \cite{carlos}
the positions of peaks $\alpha_L$'s (in which $\chi$ presents a peak) 
as well the peaks also scale with $L^{-2}$ and $L^{2}$ (panels $(d)$),
respectively. From this scaling behavior,  
we obtain the values $\alpha_0=0.5027(2)$
and $0.7844(2)$ for $D=0.1$ and $D=0.9$, respectively, 
in agreement with above estimates (from the time decay of $\rho$).

Thus, in similarity with MFT, numerical simulations
suggest that the diffusion does not change the order of phase transition,
although clusters become less compact as $D$ increases. 
Similar conclusions are found for extremely 
low diffusion strengths. For example, for the model A with
$D=0.01$ and $0.05$, all above features  are verified at  $\alpha \sim 0.215$
and $0.239$, respectively.
Moreover,  results of both models contrast partially with those obtained by 
Villa-Martin et al. \cite{munoz214} in the regime of low diffusion rates,
although  they
agree qualitatively in the limit of  intermediate and large diffusion regimes. 
A possible explanation for such differences is presented as follows:
As shown in Ref. \cite{munoz214}, a coarse-grained description
of a  discontinuous absorbing transition is the differential equation 
$\partial_{t}\rho=-\alpha\rho-b\rho^2-c\rho^3 +D\nabla^{2}\rho+
\eta(x,t)$, where $D\nabla^2\rho$ corresponds to the diffusion term
and $\eta(x,t)$ is the (white) noise. The parameter $b$
is responsible for the density discontinuity, since
their signs  $b<0$ $(>0)$ provide  two (one)
stable solutions. On the other hand, the parameter 
$c$ is required to be positive ($c>0$) for ensuring finite densities.
The deterministic part of above equation ($\alpha \rho-
b \rho^2- c\rho^3$) can be obtained for example
by taking a fully connected lattice with the second Sch\"ogl 
transition rates $W^{-}(\rho \rightarrow 
\rho-1/L^2)=\alpha\rho$ and $W^{+}(\rho \rightarrow 
\rho+1/L^2)=\rho^2(1-\rho)$, where at each time instant the system
density $\rho$ changes by a factor $\pm 1/L^2$. A similar reasoning can be 
extended for the present
studied models, but with different  $W^+$'s. In particular, they
 read $W^{+}(\rho \rightarrow 
\rho+1/L^2)=\rho^2(1-\rho)(3-3\rho+\rho^2)$ 
and $W^{+}(\rho \rightarrow \rho+1/L^2)=\rho^2(1-\rho)(6-8\rho+3\rho^2)$
for the models A and B, respectively and 
 leading to the (deterministic) terms 
$\partial_{t}\rho=-\alpha\rho+3\rho^2-6\rho^3+4\rho^4-\rho^5$ (model A) 
and $\partial_{t}\rho=-\alpha\rho+6\rho^2-14\rho^3+11\rho^4-3\rho^5$ (model B). 
Thus,   different coarse grained descriptions can explain
the difference between results in the regime of low
diffusion rates. 
It is worth remarking that  further studies are still
required to confirm above points.

Despite the similarities between models A and B, some differences
are clearly observed. In particular,
as predicted by the MFT, compact clusters are somewhat  
less dense   for  model B than for  model A. Also, the dependence
between $D$ and $\alpha$ in both models is somewhat different.

\begin{figure}[h]
\centering
\includegraphics[scale=0.4]{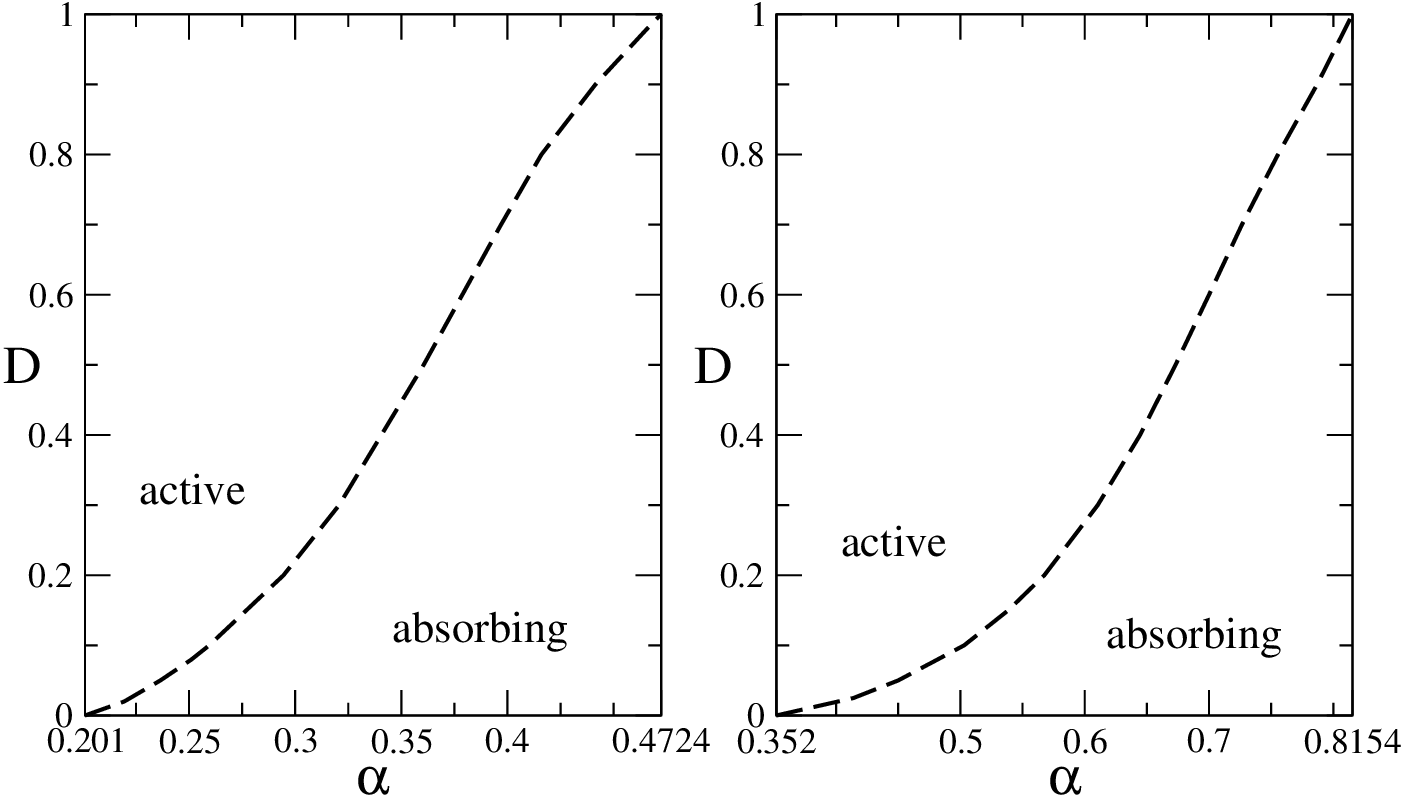}
\caption{The phase diagram 
in the plane  $D-\alpha$ obtained
from MC simulations for the models A (left) and
B (right). Dashed line
denotes discontinuous phase transitions.}
\label{fig5}
\end{figure}

Extending the aforementioned analysis for distinct values of $D$, 
we obtain the phase diagram shown in Fig. \ref{fig5}. 
For both models, in the limit $D \rightarrow 1$ 
the transition points approach their values predicted by the MFT.

\section{Conclusions}
To sum up, we investigated the influence of  diffusion
in  two simple models presenting
discontinuous absorbing phase transitions. They
have been studied via mean-field analysis and
distinct numerical simulations. All results suggest that
 transitions are discontinuous irrespective 
the  strength of the diffusion, in
contrast to recent findings in which
limited diffusion induces a critical transition \cite{munoz214}.
Thus our results indicate not only
an additional feature of diffusion, but also the possible 
differences between lattice model and  Langevin approaches. 
It is worth emphasizing that the study
of other models variants (taking into account the inclusion of distinct annihilation rates) are
required for checking whether the disagreement between
approaches is also verified in other cases.
Other remarkable result is that
all transitions presented a  finite-size scaling (with the system volume)
 similar to the discontinuous transitions  studied in Ref. \cite{carlos}. 
Although further investigations are still required, the obtained results  
reinforces the possibility
of a general scaling for nonequilibrium phase transitions.

\section*{ACKNOWLEDGMENT}
The authors wish to thank Brazilian scientific agencies CNPq, INCT-FCx for the financial
support.

\end{document}